\documentclass[aps,floats,epsf,twocolumn]{revtex4}
\usepackage{graphics}

\begin{document}

\title{Quantum fluctuations, pseudogap, and the $T=0$ 
superfluid density in strongly correlated d-wave superconductors}

\author{Igor F. Herbut}

\affiliation{Department of Physics, Simon Fraser University, 
Burnaby, British Columbia, Canada V5A 1S6 }

\begin{abstract}
I study the effect of Coulomb interaction on superconducting
order in a d-wave lattice superconductor at $T=0$ by  
considering the superconducting saddle point in the two
dimensional t-J-U model with a repulsion $U$. The theory of low-energy
superconducting phase fluctuations around this saddle point is derived 
in terms of the effective hard-core bosons (representing the density
of spin-up electrons and the phase of the order parameter),
interacting with the fluctuating density
of spin-down electrons. Whereas the saddle-point value
of the superconducting gap is found to
continuously increase towards half filling, the phase stiffness  
at $T=0$ has a maximum, and then decreases with further
underdoping. Right at half filling the phase stiffness vanishes for large
$U$. This argues that the pseudogap phenomenon of the type observed in
cuprates is in principle possible without a development of any competing
order, purely as a result of growing correlations in the superconducting
state. Implications for the finite temperature superconducting transition
and the effects of static disorder are discussed qualitatively. 
\end{abstract}
\maketitle

\section{Introduction}

 Pseudogap phenomenon has become one of the hallmarks of high temperature
 superconductivity \cite{timusk}: while the suppression of the
 single particle density of states is observed in many quantities at
 a high temperature $T^*$, which appears to {\it increase} towards
 half filling, the
 superconducting transition temperature $T_c$, together with the $T=0$
 superfluid density, at the same time continuously approaches zero
 \cite{uemura}. Being in dramatic contrast to the well understood
  behavior of the standard BCS superconductors \cite{schrieffer},
 this puzzling phenomenon has prompted
 different explanations. These may be grouped into
 at least two conceptually separate camps. The first group postulates 
 development of a second order parameter in the underdoped region, which
 competes with superconductivity and suppresses its
 transition temperature \cite{varma},\cite{chakravarty}, \cite{zhang},
 \cite{lee}, \cite{sachdev}, \cite{tallon}, \cite{tremblay}.
 The second interprets
 $T^*$ as a crossover temperature at which the Cooper pairs, loosely
 speaking, begin to form, and the
 lower $T_c$ as the point where the phase coherence finally sets in
 \cite{doniach},
 \cite{emery}, \cite{loktev}. The suppression of $T_c$ is then 
 attributed to
 the gradual loss of carriers near the insulating state at half filling.
 There is a significant amount of possible empirical support for the latter
 point of view: d-wave symmetry of the pseudogap \cite{shen},
 specific heat measurements \cite{olson}, heat
 transport \cite{sutherland}, microwave conductivity \cite{corson},
 and the Nernst effect  \cite{ong} may all be understood
 as directly or indirectly supporting the proposed 
 superconducting origin of the pseudogap temperature.
Some of the same measurements, however, may also be understood
within a competing theory, and the physics of underdoped high
temperature superconductors at present still remains controversial. 

In this paper it is argued that the pseudogap phenomenon may
in principle arise purely from strong Coulomb interactions in a
d-wave superconductor and sufficiently near half filling, {\it without}
any competing order. While a competing order is possible, and indeed
in some materials may even be likely in the underdoped region,
it seems not to be required simply by the existence of pseudogap.
Furthermore, this is shown for a superconductor
with only a {\it weak} attraction in the d-wave channel, so the mechanism
behind the pseudogap considered here is different than in the 
real-space pairing approaches extensively discussed in the literature
\cite{randeria}, \cite{loktev}. Deriving from Coulomb repulsion,  
it is similar in spirit, although still different in detail, to
the one in the RVB theory \cite{anderson}, \cite{rice}, \cite{paramekanti}. 

We consider what is probably the simplest 
model which contains the relevant physics, the t-J-U model
in Eq. (1), with the standard exchange term rewritten as
the pairing interaction and with a strong
 repulsion, and study the evolution of the $T=0$ superconducting
gap and the phase stiffness with doping. Independently of its
magnitude, the repulsion does not affect
the dependence of the mean-field d-wave superconducting (dSC) gap on the number
of particles, as it only shifts the (unphysical) bare value of the chemical
potential. As a result, the saddle point value of the superconducting gap
(Figure 1) continuously increases towards
half filling, as found in other similar calculations \cite{kotliar}.
It is intuitively clear however, that for a large repulsion $U(x)$
the superconducting order should eventually 
be weakened near half filling, since the Cooper pairs, 
although formed, will have little space left to move. It is less
obvious in which exact fashion should this occur; for example,
should the suppression be gradual or abrupt \cite{kivelson}, or what
its effect on quasiparticles should be \cite{ft}, \cite{qed3}. It is possible to
make the physics behind this `jamming' effect quite explicit by introducing
certain collective
coordinates to describe the (quantum) phase fluctuations around the dSC saddle point:
{\it two} densities of spin-up and spin-down electrons, and the phase of
the superconducting order parameter.  The  
crucial feature of our representation is that {\it only one} of the
electron densities (say of spin-up) becomes the conjugate variable
to the superconducting phase. This enables one to understand quantum
($T=0$) phase fluctuations around the dSC saddle point within the 
theory of interacting bosons (representing jointly the density of
spin-up electrons and the order parameter phase) moving in the fluctuating
and interacting background provided by the density of  the remaining 
spin-down electrons. A self-consistent calculation of the $T=0$
phase stiffness in such an effective  bosonic system
yields then generically a non-monotonic behavior with doping (Figure 2).
For $U$ large enough the phase stiffness at half filling vanishes, while
the superconducting gap remains perfectly finite, and in fact large.

 The superconducting $T_c$ is then non-monotonic because it is
 determined by the smaller of the two characteristic temperatures
 \cite{emery}, one set by
 the superconducting gap, and the other by $T=0$ superfluid density.
 It is not the bare superfluid density 
 that vanishes near half-filling, however, but only the physical
 (renormalized) one that becomes reduced by 
 quantum fluctuations. The bare one, indeed, only increases
 towards half filling, since in a weakly coupled superconductor
 it is proportional to the electron density. Furthermore,
 in contrast to $T_c$ $T=0$ superfluid density should
 continue to grow as doping is increased through the optimal
 doping, as observed in number of experiments \cite{schneider}. The
 characteristic temperature dependence of the superfluid density under the 
 conditions of a large d-wave gap and a low $T_c$ 
 is discussed in a separate publication \cite{case}. 

 In the previous work on the d-wave quasiparticles \cite{qed3}
 it was assumed that
 underdoping, known from experience to be detrimental to superfluid density,
 could be viewed as inducing quantum phase fluctuations into an
 otherwise standard d-wave BCS-like superconducting ground state.
 Present results support that hypothesis. It was also shown that
such quantum phase fluctuations lead generically to spin response
which agrees in many details with the
 observations in underdoped Y-Ba-Cu-O (YBCO) \cite{herbutlee}. Together with the
 present results, this lends additional credence to the idea that
 the underdoped cuprates may be understood as  strongly (quantum) phase
 fluctuating d-wave superconductors. The effect of low phase stiffness
 on charge degrees of freedom will be discussed in a future publication
 \cite{igorcharge}.

  The paper is organized as follows. In sec. II the $t-J-U$ Hamiltonian
is introduced, and the superconducting saddle point is discussed in the
 functional integral formulation. The $T=0$ theory for the phase fluctuations
 around the saddle point is derived in sec. III. The effective bosonic
 theory is presented in sec. IV, together with the self-consistent
 calculation of the $T=0$ superfluid density. The discussion of the
 results in relation to the physics of cuprates is given in sec. V.

\section{Model and the superconducting saddle point}

We begin by defining the t-J-U Hamiltonian. Similar models have been often
invoked in the past to study strongly correlated electrons \cite{plekhanov},
 \cite{martin}, \cite{fczhang}. Consider the imaginary time
quantum mechanical action $S=\int_0 ^\beta d\tau L(\tau)$, $\beta=1/T$, 
and $L(\tau)=\sum_{x,\sigma} c^\dagger _{\sigma}(x,\tau)
  (\partial_\tau -\mu) c_{\sigma} (x,\tau)
  + H(\tau)$, with $c_{\sigma}(x,\tau)$ being the standard Grassman electronic
  variables, and the Hamiltonian 
  \begin{eqnarray}
  H(\tau) = -t \sum_{\langle x,x'\rangle, \sigma=\pm }
  c^\dagger _\sigma (x,\tau ) c_\sigma (x',\tau) + \\ \nonumber 
  \sum_{x,x'} (n_+ (x,\tau) +n_- (x,\tau)) \frac{U(x-x')}{2} 
   (n_+ (x',\tau) +n_-(x',\tau))  \\ \nonumber
  + J \sum_{\langle x,x'\rangle }
  [ \vec{S}(x,\tau) \cdot
  \vec{S}(x',\tau) - \frac{1}{4} \sum_{\sigma, \sigma'}
  n_\sigma (x,\tau) n_{\sigma'} (x',\tau) ]. 
  \end{eqnarray}
  Here, $\vec{S}(x,\tau) = (1/2) \sum_{\alpha,\beta}
  c^\dagger _\alpha (x,\tau) \vec{\sigma}_{\alpha \beta}
  c_\beta (x,\tau)$ and $n _\sigma (x,\tau) = c^\dagger _\sigma (x,\tau)
  c_\sigma (x,\tau)$
  are the standard electron spin and particle densities, $U(x),J >0$, and
  $x$ labels sites of a two dimensional quadratic lattice. When
  $U(x)=U \delta_{x,x'}$ with $U\rightarrow
  \infty$ the model becomes equivalent to the standard t-J model as
  derivable from the underlying Hubbard model, since
  double occupancy then becomes
  completely suppressed. Here we will relax this
  condition, and consider a general, and finite repulsive interaction $U(x)$.
  We will however, be particularly
  interested in the regime of parameters $U(x) \gg t>J$, which  corresponds to
  a {\it weakly coupled}, but a {\it strongly correlated} superconductor.

\begin{figure}[t]
{\centering\resizebox*{80mm}{!}{\includegraphics{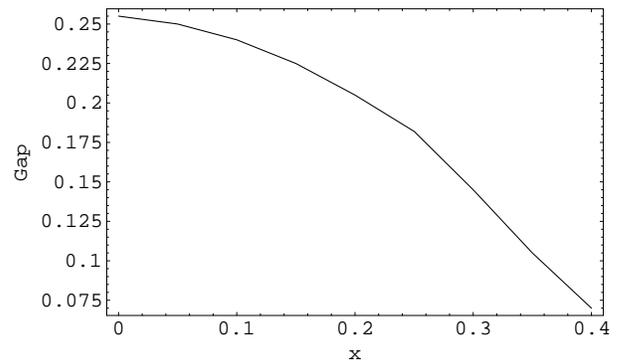}}\par}
\caption[]{The saddle point of the gap $\Delta_0 /t$ vs. doping $x=1-n$,
from Eqs. (4) and (5).}
\label{gap}
\end{figure}

It is presently controversial \cite{sorella}, \cite{pryadko}
whether the canonical 
t-J model indeed supports the d-wave superconducting ground state.
Although interesting, this issue will be of little
concern to us here. Our philosophy \cite{balents}, \cite{ft}, \cite{qed3}
 shared by number
of recent works \cite{paramekanti}, \cite{laughlin}, is to start from
the experimental fact that cuprates 
are d-wave superconductors at least for a range of dopings \cite{tsui},
and take
that as the point of departure for further exploration of the phase
diagram. We will therefore regard the Hamiltonian (1)
merely as a useful tool for introducing correlations
into the {\it postulated} superconducting ground state.

 The last ($\sim J$) term in the Hamiltonian
can be rewritten to explicate its pairing nature as
  \begin{equation}
  H_J= -\frac{J}{2} \sum_{\langle x, x'\rangle} B^\dagger (x,x',\tau')
  B(x,x',\tau),
  \end{equation}
with $B(x,x') = c_- (x',\tau) c_+ (x,\tau) - c_+ (x',\tau) c_- (x,\tau)$
annihilating a singlet on a pair of neighboring sites. In the momentum space, $H_J$ 
is a sum of the d-wave and the extended s-wave pairing terms
\cite{paramekanti}. In what follows we
consider a purely d-wave  saddle point, and neglect completely
a possible s-wave component. Being small at the Fermi surface near half
filing, the amplitude of the s-wave order parameter is expected to be
completely suppressed by the dominate d-wave component \cite{wallington}.

We start by  decoupling the $H_J$ term using a complex Hubbard-Stratonovich
field $\Delta(x,x',\tau)$, and by introducing two density variables
$\rho_\sigma (x,\tau)$, $\sigma=\pm$ as
\begin{eqnarray}
L(\tau) = \sum_{\langle x,x' \rangle, \sigma}  c^\dagger _\sigma (x,\tau)
[ (\partial_\tau - \mu + i j_\sigma (x))\delta_{x,x'}-t]c_\sigma (x',\tau) \\ \nonumber
-i\sum_{x,\sigma} j_\sigma (x,\tau) \rho_\sigma (x,\tau) +
\frac{1}{2} \sum_{x,x'} \rho (x,\tau) U(x-x') \rho (x' ,\tau) \\ \nonumber
+ \sum _{\langle x,x' \rangle }
[\frac{2}{J} |\Delta(x,x',\tau)|^2 +
\Delta (x,x',\tau) B^\dagger (x,x',\tau) + H.c. ] 
\end{eqnarray}
The variables $j_\sigma (x,\tau)$ are the Lagrange multipliers that enforce the
constraints $\rho_\sigma (x,\tau) = n_\sigma (x,\tau)$ \cite{depalo},
and $\rho(x,\tau)=\rho_+(x,\tau)+\rho_-(x,\tau)$. The d-wave superconducting
saddle point is as usual given by $\Delta(x,x',\tau) = +\Delta_0$
if $x' = x+\hat{1}$ and
$\Delta(x,x',\tau) = - \Delta_0$ if $x'= x+\hat{2}$,
where $\hat{1}$ and $\hat{2}$ are the unit lattice vectors,
and with the amplitude $\Delta_0
\neq 0 $ at $T=0$ being determined by the standard BCS gap equation
\begin{equation}
1=J\int \frac{d^2 \vec{k}}{(2\pi)^2} \frac{\phi_d ^2 (\vec{k})}
{\sqrt{( e(\vec{k})-\tilde{\mu})^2 + |\Delta_0|^2 \phi_d ^2 (\vec{k}) }}, 
\end{equation}
together with the number equation 
\begin{equation}
n= \int \frac{d^2 \vec{k}}{(2\pi)^2}  [ 1- \frac{ e(\vec{k}) -\tilde{\mu} } 
{\sqrt{( e(\vec{k})-\tilde{\mu})^2 + |\Delta_0|^2 \phi_d ^2 (\vec{k}) }}].
\end{equation}
$n$ represents the number of electrons per site, related to doping
$x$ as $n=1-x$. Here
$e(\vec{k}) = -2 t (\cos(k_x) + \cos(k_y))$, $\phi_d (\vec{k})=\cos(k_x) -
\cos(k_y)$,
and $\tilde{\mu}= \mu - nU(q=0)$. We assumed a {\it non-magnetic} saddle point
at which $\rho_+=\rho_- =n/2$, and $j_+ = j_- = -in U(q=0)$
\cite{remark}. Our goal here is to consider quantum fluctuations
resulting purely from the electron repulsion, i. e. {\it without}
any explicit competing order. This is not to say that some other
saddle point may not become stable as the electron density or the interaction
$U$ is varied; indeed, in the particular t-J-U model
this would be expected for large $U$. Our concern here, however,
is not the phase diagram of the specific t-J-U model \cite{martin},
but the broader issue of the evolution of the postulated
 superconducting ground state in presence of strong repulsion.

An example of a numerical solution of the saddle-point Eqs. (4)-(5)
 is presented in Figure 1, for parameters $t=150 meV$ and $J=100 meV$. 
Even with such a relatively large $J$ the
superconductor is effectively still in the weak-coupling regime,
as evidenced by the chemical
potential, for example, changing very little by the opening of the
superconducting gap. In the rest of the paper we will therefore be
concerned exclusively with such a weakly-coupled superconducting state.
 Figure 1 should of course be understood only as an illustration,
as a more realistic calculation should include at least a sizable 
next-nearest-neighbor hopping $t'$. It does demonstrate, however, two
expected and generic
features: a) the gap at a given electron
density is unaffected by the repulsion $U(x)$, which only
shifts the bare value of the chemical potential $\mu$, b) the saddle-point
value of the gap is a
{\it monotonically increasing} function of the electron density. The mean-field analysis 
by itself would therefore suggest an increasing superconducting $T_c$ towards
half filling, in blatant contradiction to the generic behavior of
underdoped cuprates. This conclusion will be overturned however by the
inclusion of quantum phase fluctuations introduced by the large 
repulsion, as discussed next.

\section{Quantum fluctuations}

   To this end, consider the fluctuations of the phase of the
order parameter, $\Delta (x,x+\hat{a},\tau) = (-)^a \Delta_0
e^{-i \theta (x,\tau)}$, where $\hat{a}= \hat{1},\hat{2}$.
To reduce the algebraic complexity we will neglect the fluctuations of the
amplitude of the order parameter, which should be justified at
$T=0$. We are also  assuming a {\it single} phase
$\theta(x,\tau)$ for both links $(x,x+\hat{i})$ and $(x,x+\hat{j})$ emanating from 
the site $x$. In doing so we are
neglecting the fluctuations towards the possible s-wave component of the
superconducting order, in accord with the expectation that
it is suppressed by the dominate d-wave order \cite{wallington}. 

As usual, at this stage one would like
to integrate out fermions, to be left with an effective action for the
collective variables.  To perform this step, we first introduce new
Grassman variables as
$c^\dagger _+ (x,\tau)=e^{i\theta(x,\tau)} a^\dagger _+ (x,\tau)$, 
$c_+ (x,\tau)=e^{-i\theta(x,\tau)} a_+ (x,\tau)$, 
$c^\dagger _- (x,\tau) = a^\dagger _- (x,\tau)$, and
$c_- (x,\tau) = a_- (x,\tau)$. Note that in contrast to  
previous works \cite{mandal}, \cite{benfatto},
the {\it entire} superconducting phase here is 'absorbed' into 
particles of a single spin-projection, arbitrarily chosen to be up
\cite{anders}.
This guarantees that for {\it all} possible configurations of the
phase, including vortices, new Grassman variables $a$ and $a^\dagger$
are single valued and satisfy the standard fermionic boundary conditions at
$\tau=\beta$. Although the above change of variables is not unique
in accomplishing this it probably is the simplest. 

Next, as usual, introduce the deviations from the saddle-point values
of the fields as  $i j_\sigma (x,\tau)= \delta j_\sigma (x,\tau)+ n U(q=0)$,
$\rho_\sigma (x,\tau)= \delta \rho_\sigma (x,\tau)+ n/2$, $\sigma=\pm$,
and then shift $\delta j_+ (x,\tau)- i \partial_\tau \theta (x,\tau)
\rightarrow \delta j_+ (x,\tau)$.
The Lagrangian then becomes $L(\tau)=L_a (\tau) + L_\rho(\tau)$, with 
\begin{eqnarray}
L_a (\tau)= \sum_{\langle x,x' \rangle, \sigma}  a^\dagger _\sigma (x,\tau)
[ (\partial_\tau - \tilde{\mu} + \delta j_\sigma (x,\tau))\delta_{x,x'} \\ \nonumber
-t \delta_{\sigma, -}  - t e^{i(\theta(x,\tau)-\theta(x',\tau))}
\delta_{\sigma,+}]a_\sigma (x',\tau)+ \\ \nonumber 
\sum _{\langle x,x' \rangle } \pm \Delta_0  
[ a^\dagger _+ (x,\tau) a^\dagger _- (x',\tau) -   e^{-i(\theta(x,\tau)-
\theta(x',\tau))} \\ \nonumber
a^\dagger _- (x,\tau)  a^\dagger _+ (x',\tau)  + H.c.],
\end{eqnarray}
\begin{eqnarray}
L_\rho(\tau)=  -\sum_{x,\sigma} \delta j_\sigma (x,\tau)\rho_\sigma (x,\tau) +
\\ \nonumber
\frac{1}{2} \sum_{x,y} \delta \rho (x,\tau) U(x-y)
\delta \rho (y,\tau)
- i \sum_x \rho_+ (x,\tau) \partial_\tau \theta(x,\tau). 
\end{eqnarray}
The shift of one of the Lagrange multipliers serves to isolate the
imaginary-time derivative of the phase into the last term in $L_\rho$,
and thus promote the density of spin-up electrons $\rho_+$ to the status of 
{\it conjugate variable} to the superconducting phase. It is  important
to note that it is not $(\rho_+(x,\tau) + \rho_- (x,\tau) )/2 $ that
plays the role of the conjugate variable, as it would have been obtained
if the two projections of spin were treated equally, and half of the
superconducting phase absorbed into each species.  That, often invoked
transformation is allowed only for the trivial boundary condition
 $\theta(x,\beta)=\theta(x,0)+ 2\pi n$, with $n=0$,
 and with the vortices in the phase being forbidden. For this trivial 
 boundary condition the last term in Eq. 7 becomes independent
 of the average particle density, and any effect of the proximity to
 commensurate density, which is our main subject, is lost. To have any
 commensuration effects on the superfluid response it is thus paramount to
 allow for the non-trivial (in the above sense) phase configurations.

In principle the fermions can now be integrated out. The resulting
action will be a functional of the phase $\theta(x,\tau)$ and the two
densities $\rho_\pm (x,\tau)$ of spin-up and spin-down electrons. The fact
that $\rho_+$ and $\theta$ form a pair of conjugate variables suggests a
particular change of variables, so that the action becomes 
a functional of the {\it bosonic} variable $\Psi=\sqrt{\rho_+}
e^{i\theta}$, $\Psi^*$, and $\rho_-$. This is the central
idea of this paper. In the rest we will try to approximately determine
the functional $S [\Psi,\Psi^*,\rho_-]$, and then use it
to compute the stiffness for the superconducting
phase fluctuations at $T=0$. Alternatively, the lattice action in
Eqs. 6 and 7 can be studied numerically. 

In practice however, the integration over fermions is performable analytically
only for small phase gradients and a small $\delta j_\sigma$. After a
straightforward but a rather involved algebra one finds
$ S= S_F + \int_0 ^\beta d\tau L_\rho (\tau) $,
 with the contribution from the fermion determinant
\begin{eqnarray}
S_F= \frac{1}{2} \int \frac{ d^2 \vec{q} d\nu}{(2\pi)^3}
 R^\dagger (\vec{q},\nu)
 \hat{M}(\vec{q},\nu) R(\vec{q},\nu) + \\ \nonumber 
 \frac{n}{2}\int_0 ^\beta d\tau \sum_{x,\sigma} \delta j_\sigma (x,\tau) 
 + O(R^4),
\end{eqnarray}
where $ R^\dagger (\vec{q},\nu) = (\delta j_+ (\vec{q},\nu),
-i \theta(\vec{q},\nu),
\delta j_- (\vec{q},\nu) )$, and the $3\times 3$ matrix $\hat{M}$
is symmetric with the elements
\begin{equation}
M_{11}(\vec{q},\nu)= M_{33}(\vec{q},\nu)=
\int \frac{ d^2 \vec{q} d\omega}{(2\pi)^3}
 G(\vec{k},\omega) G(\vec{k}+
\vec{q}, \omega+\nu), 
\end{equation}
\begin{equation}
M_{13}(\vec{q},\nu)= - \int \frac{ d^2 \vec{q} d\omega}{(2\pi)^3}
 F(\vec{k},\omega) F(\vec{k}+
\vec{q}, \omega+\nu), 
\end{equation}
\begin{eqnarray}
M_{12}(\vec{q},\nu)= \int \frac{ d^2 \vec{q} d\omega}{(2\pi)^3}
G(\vec{k}+\vec{q},\omega+\nu)
[ ( e(\vec{k}+\vec{q})\\ \nonumber
-e(\vec{k}) )  G(\vec{k},\omega)
+ ( \Delta (\vec{k}+\vec{q})-\Delta(\vec{k}) )F(\vec{k},\omega)]  \\ \nonumber
+(\Delta(\vec{k}+\vec{q})-\Delta(\vec{k}) ) F(\vec{k}+\vec{q},\omega+\nu)
G(\vec{k},\omega) ,
\end{eqnarray}
\begin{eqnarray}
M_{23}(\vec{q},\nu)=
\int \frac{ d^2 \vec{q} d\omega}{(2\pi)^3}
F(\vec{k}+\vec{q},\omega+\nu)
[ -( e(\vec{k}+\vec{q})\\ \nonumber
-e(\vec{k}) )  F(\vec{k},\omega)
+( \Delta(\vec{k}+\vec{q})-\Delta (\vec{k}) )G(-\vec{k},-\omega) ] 
\\ \nonumber
+(\Delta (\vec{k}+\vec{q})-\Delta (\vec{k}) ) G(-\vec{k}-\vec{q},-\omega-\nu)
F(\vec{k},\omega) ,
\end{eqnarray}
and  finally
\begin{eqnarray}
M_{22} (\vec{q},\nu)=  \\ \nonumber
\int \frac{ d^2 \vec{q} d\omega}{(2\pi)^3}
[ e(\vec{k}+\vec{q})-e(\vec{k}) ]^2  G(\vec{k},\omega) G(\vec{k}+\vec{q},
\omega+\nu )\\ \nonumber 
+2 [ e(\vec{k}+\vec{q})-e(\vec{k}) ] 
( \Delta(\vec{k}+\vec{q})-\Delta(\vec{k}) ) [G(\vec{k}+\vec{q},\omega+\nu )
\\ \nonumber
F(\vec{k},\omega) + G(\vec{k},\omega) F(\vec{k}+\vec{q},\omega+\nu ) ] 
+2 [ \Delta (\vec{k}+\vec{q})-\Delta (\vec{k}) ] ^2 \\ \nonumber
[ F(\vec{k},\omega) F(\vec{k}+\vec{q},\omega+\nu )  
-G(-\vec{k},-\omega) G(\vec{k}+\vec{q},\omega+\nu ) ]   \\ \nonumber
+ 2  [ e(\vec{k}+\vec{q})-e(\vec{k}) ] G(k,\omega)  \\ \nonumber
- 2 [ \Delta(-\vec{k}+\vec{q})-\Delta(-\vec{k}) ] F(k,\omega),  
\end{eqnarray}
with $G(\vec{k},\omega) = (-i \omega - e(\vec{k}) + \tilde{\mu})/[\omega^2 +
\Delta^2 (\vec{k}) + (e(\vec{k})-\tilde{\mu})^2] $, and
$F(\vec{k},\omega) = \Delta(\vec{k},\omega) /
[\omega^2 +\Delta^2 (\vec{k}) + (e(\vec{k})-\tilde{\mu})^2] $,
and $\Delta(\vec{k}) = \Delta_0 \phi_d (\vec{k})$.  Note that Eqs. (9)-(13)
are similar, but not identical to those that can be found
in \cite{benfatto}, for example. This 
is due to a different change of fermionic variables employed
there which treats spin-up and spin-down symmetrically, as discussed
earlier. In particular, we have $M_{12}\neq M_{23}$, and as a result the two
densities are in principle coupled differently to the gradients of the phase.
This serves to compensate for the asymmetry between $\rho_+$ and $\rho_-$ in
$L_\rho$, and to insure that if solved exactly, the theory would lead
to equal average densities of up and down particles.

 The elements of $\hat{M}(\vec{q},\nu)$ can be easily evaluated
 for a weakly coupled superconductor with $\Delta_0 \ll t$.
 Assuming a spherical Fermi
 surface for convenience, we find $M_{11}(\vec{q},\nu) = -{\cal N}/4 +
 O(q^2, \nu^2)$, $M_{13} (\vec{q},\nu) = -{\cal N}/4 + O(q^2, \nu^2)$,
 $M_{12} (\vec{q},\nu)= M_{23} (\vec{q},\nu)= O(q^2)$,
 and $ M_{22}(\vec{q},\nu) =
 (e_F/4\pi)(1+ O((\Delta_0/e_F)^2)) q^2 + O( q^4, q^2 \nu^2)$,
 where ${\cal N}$ is the density of states at the Fermi level $e_F$   
 (${\cal N} = m^* /\pi$ for the spherical dispersion
 $e(k)=k^2 /2m^*$, with $m^*=1 /(2t a^2) $ being an effective electron mass,
 and $a$ the lattice spacing).

   At long distances ($q,\nu \rightarrow 0$) one can therefore neglect
 the matrix elements $M_{12}$ and $M_{23}$  that couple 
the fluctuating phase to the Lagrange multipliers. The integration over
$\delta j_\pm$ then gives the action in terms of the  densities and
the phase to be 
\begin{eqnarray}
S= \int \frac{ d^2 \vec{q} d\nu }{(2\pi)^3}
\{ \frac{e_F}{8\pi} q^2 \theta(\vec{q},\nu) \theta(-\vec{q},-\nu)+\\ \nonumber
\frac{M_{11}(\vec{q},\nu)}{2( M_{13}^2 (\vec{q},\nu)- M_{11}^2
(\vec{q},\nu) ) }  
(\delta \rho_+ ^2 (\vec{q},\nu)  +\delta \rho_- ^2 (\vec{q},\nu) )- \\ \nonumber
\frac{M_{13}(\vec{q},\nu)}{M_{13}^2 (\vec{q},\nu)- M_{11}^2 (\vec{q},\nu)}
 \delta\rho_+ (\vec{q},\nu)\delta\rho_- (-\vec{q},-\nu)+ \\ \nonumber 
\frac{1}{2} U(q) \delta \rho(\vec{q},\nu) \delta \rho(-\vec{q},-\nu)
+ \nu \theta(\vec{q},\nu) \rho_+ (-\vec{q},-\nu) \},
\end{eqnarray}
where we retained only the leading terms in the expansion of $S$ in powers of
phases, densities, and their derivatives. The bare phase
stiffness $\sim e_F= 2\pi n t $ for weak coupling, and in general will be 
proportional to the total kinetic energy of electrons in the
superconducting state.

The integration over fermions has introduced additional
quasi-interactions between the electron densities in Eq. 14, on top of the
repulsion $U(q)$:
\begin{eqnarray}
V_{++} (\vec{q},\nu)= V_{--}(\vec{q},\nu)=
\frac{M_{11}(\vec{q},\nu)}
{( M_{13}^2 (\vec{q},\nu)- M_{11}^2 (\vec{q},\nu) ) } \\ \nonumber 
\approx \frac{1}{2 (M_{13}(\vec{q},\nu) - M_{11} (\vec{q},\nu)) }, 
\end{eqnarray} 
and
\begin{eqnarray}
V_{-+} (\vec{q},\nu)=
-\frac{M_{13}(\vec{q},\nu)}{ M_{13}^2 (\vec{q},\nu)- M_{11}^2 (\vec{q},\nu) )}
\\ \nonumber  
\approx - \frac{1}{2 ( M_{13}(\vec{q},\nu) - M_{11} (\vec{q},\nu) ) }. 
\end{eqnarray}
The quasi-interaction is repulsive between the same and attractive
between the opposite spin densities. The approximate expressions in Eqs.
(15) and (16) are valid
at small $q$. Here we utilized the fact that (for both d-wave and
s-wave superconducting state) at $T=0$
\begin{equation}
M_{13}(0,\nu) = M_{11}(0,\nu), 
\end{equation}
exactly. This is easily established using the definitions of the
matrix elements and performing the frequency integral.
Physically, this simply expresses the fact that the uniform spin
susceptibility in a superconductor with singlet pairing
vanishes at $T=0$. This is probably most easily seen by 'rotating'
our density variables
into total density and magnetization, as $\rho=\rho_+ + 
\rho_-$, $m=\rho_+ -  \rho_-$,
and then reading off the [random-phase approximation (RPA)] spin and
charge susceptibility in the superconductor from Eq. (14):
\begin{equation}
\chi_s ^{-1} (\vec{q},\nu) = \frac{1}{2(M_{13}(\vec{q},\nu)
- M_{11}(\vec{q},\nu))}, 
\end{equation}
\begin{equation}
\chi_c ^{-1} (\vec{q},\nu) = U(q) - \frac{1}{2(M_{13}(\vec{q},\nu)
+ M_{11}(\vec{q},\nu)) }. 
  \end{equation}
Eq. (17) therefore implies a vanishing uniform
spin susceptibility and a finite compressibility (for short-range interaction)
in a singlet superconductor
at $T=0$. Note also that for non-interacting electrons ($U=J=0$) one recovers
$\chi_s=\chi_c = -2 M_{11}(q,\nu)$, which is then the familiar Lindhard
response.

We may gain further insight into the nature of quasi-interactions 
$V_{++}$ and $V_{+-}$ by computing them at small $q$. In the first
approximation one may neglect the anisotropy of the d-wave gap and perform
the calculation for the simpler s-wave superconductor. The result is 
\begin{equation}
M_{13}(\vec{q},\nu) - M_{11}(\vec{q},\nu) \approx
\frac{e_F}{6 \pi \Delta_0 ^2} q^2 (1+ O(\nu^2)).
\end{equation}
We checked numerically that the result for d-wave superconductor is
essentially the same, except for the anisotropy in the coefficient
of the $q^2$-term. Taking the d-wave nature of the gap into account 
yields a maximum of $M_{13}-M_{11}$ at a finite $\vec{q}$ that connects the
gap nodes, so that the Stoner instability of a metal in a dSC becomes
replaced by an analogous RPA instability towards the spin density wave
\cite{lu}. For the present purposes it suffices to note that the
quasi-interactions induced by the fermion integration are long-ranged,
but {\it weak}, for a weakly coupled superconductor, i. e.
$V_{\sigma \sigma'} \sim \Delta_0 ^2 / e_F$. In particular, this means
that the electrons with the opposite spin, although paired in
the momentum space by construction, may to a good approximation be considered
as independent in real space, and hence to be interacting primarily via
 $U(x)$.

\section{Effective bosonic Hamiltonian and the phase stiffness}

 If one would integrate out the Gaussian
 density fluctuations in Eq. 14 the remaining
 action for the phase, apart from the linear time derivative term, describes
 the standard collective mode in a weakly coupled superconductor.
 To capture the effects of nearly commensurate density on the superfluid
 response one needs to go beyond this hydrodynamic regime
 \cite{epsilon}. What is needed is
 the bosonic lattice action $S[\Psi,\Psi^*,\rho_-]$, which would correspond to
 Eqs. 6 and 7, and which at long length scales would reduce to Eq. 14.
 The simplest candidate satisfying these requirements is:
\begin{eqnarray}
L_B (\tau) = \sum_x \Psi^* (x,\tau) \partial_\tau \Psi(x,\tau) -\\ \nonumber
t_B \sum_{\langle x,x'\rangle} \Psi^* (x,\tau) \Psi (x', \tau)\\ \nonumber 
+\frac{1}{2}
\int d\tau' \sum_{x,x'} |\Psi(x,\tau)|^2 ( V_{++}(x-x',\tau-\tau') \\ \nonumber
+ U(x-x') \delta(\tau-\tau')) |\Psi(x',\tau')|^2  \\ \nonumber
+ \int d\tau' \sum_{x,x'} |\Psi(x,\tau)|^2
(U(x-x')\delta(\tau-\tau') \\ \nonumber 
+ V_{+-} (x-x',\tau-\tau') ) \rho_- (x' ,\tau') \\ \nonumber 
- \mu_B \sum_{x} (|\Psi(x,\tau)|^2 + \rho_- (x,\tau)) \\ \nonumber 
+\frac{1}{2} \int d\tau' \sum_{x,x'}\rho_- (x,\tau) (V_{--}(x-x',\tau-\tau')
\\ \nonumber
+U(x-x')\delta(\tau-\tau')) \rho_- (x',\tau'),
 \end{eqnarray}
with $\Psi (x,\tau)=\sqrt{ \rho_+ (x,\tau)} e^{i \theta(x,\tau)}$,
$t_B= e_F/(4 n \pi)$ and $\mu_B= \chi_c ^{-1} (0,0) n/2$.
The reader is invited to check that when expanded in powers
of the density and phase and their derivatives \cite{popov},
the leading order terms in $L_B$ indeed reproduce the Eq. (14).

\begin{figure}[t]
{\centering\resizebox*{80mm}{!}{\includegraphics{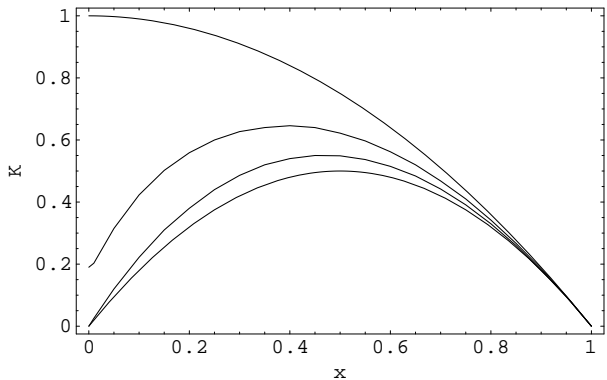}}\par}
\caption[]{The phase stiffness $K/4t_B$ in the effective bosonic theory for
phase fluctuations for $U=0$, $U/4t_b= 1.8$, $U/4t_B = 5$,
and $t_B /U = 0$ (top to bottom) vs. doping $x=1-n$. }
\label{sfdensity}
\end{figure}

The action in Eq. (21) represents bosonic particles hopping on a lattice
and interacting with a fluctuating background provided by the density
of spin-down electrons. Although the amplitude of our bosonic
field is proportional to the density of spin-up electrons only,
its phase is still the full superconducting phase and therefore the
bosons have the electromagnetic charge $2e$.  Since $e_F \sim t n$,
hopping amplitude for the bosons is $t_B = e_F/(4 n\pi )\sim t$,
and is approximately doping independent.
Since the change of fermionic variables leading to Eq. (7)
broke the spin up-down symmetry,
although the chemical potential $\mu_B$ is the same for both the bosons and
the spin-down particles in $L_B$, two densities in our approximation will not
automatically be equal as they should. One may easily 
correct this by allowing for two different chemical
potentials $\mu_+$ and $\mu_-$, and adjust them so that the average
densities of bosons and spin-down electrons become equal, as of course they
are in the full theory in Eqs. 6 and 7. Furthermore, assuming the 
Coulomb interaction $U(x)\sim 1/x$, for
a weakly coupled superconductor we may neglect the quasi-interactions
$V_{+-}$ and $V_{++}$, which are weak, and also less singular than $U(x)$ at
$x=0$ \cite{remark1}. To simplify the calculation that follows, after
neglecting the quasi-interactions 
we may further replace the realistic Coulomb interaction by
 the hard-core repulsion between the particles.
To make the calculation a little more
interesting however, we will assume the hard-core repulsion only between
the same spin particles \cite{comment}, and allow for some finite on-site
repulsion $U$ between the particles with opposite spin. The single parameter
$U$ then effectively measures the degree of correlations at the
superconducting saddle point. For $U=\infty$ then all the particles interact
 via hard-core repulsion. 

  The phase stiffness from Eq. (21) is therefore
\begin{equation}
  K= t_B |\langle \Psi(x,\tau)\rangle |^2.  
\end{equation}
Within the local approximation discussed above we may compute $K$
in the effective bosonic action in Eq. (21) by decoupling the
hopping term with a Hubbard-Stratonovich field $\Phi(x,\tau)$,
where $\langle \Phi(x,\tau)\rangle = 2 t_B \langle \Psi (x,\tau)\rangle $,
and by making yet another saddle-point approximation in the
resulting action. Assuming
that at such a saddle point both $\Phi(x,\tau)$ and $\rho_- (x,\tau)$ are
independent of the imaginary time, the 'mean-field'
bosonic free energy becomes (see Appendix)
\begin{eqnarray}
F_{B,MF}= \frac{1}{4} \sum_{x,x'} \Phi^* (x) (t_B ^{-1})_{x,x'} \Phi(x') \\ \nonumber
-T\sum_{x} \ln  Tr Exp -\frac{1}{T}
[ \frac{1}{2}\Phi(x) \hat{\Psi}^\dagger (x)
+ \frac{1}{2} \Phi^* (x) \hat{\Psi} (x)     \\ \nonumber
-\mu_+ \hat{\Psi}^\dagger (x) \hat{\Psi}(x) -\mu_- \rho_- (x)
+ U \hat{\Psi}^\dagger (x) \hat{\Psi}(x) \rho_- (x) ] ,
\end{eqnarray}
where the trace is to be evaluated over four possible states: empty site
$|0\rangle$, boson $\hat{\Psi}^\dagger (x) |0\rangle$, spin-down electron
$|- \rangle$, and boson and spin-down electron 
$\hat{\Psi}^\dagger (x) |-\rangle$. Since we assumed a static saddle point,
the temperature $T$ appears explicitly in $F_B$, and at the end we
need to impose the limit $T \rightarrow 0$.  $\Phi (x)$ takes the value that
minimizes $F_{B,MF}$, and the chemical potentials $\mu_{\pm}$ are to be tuned so
that the densities of bosons and of spin down electrons are both equal to
half the electron density:
\begin{equation}
\frac{\partial F_{B,MF} }{\partial \Phi (x)} =0, 
\end{equation}
\begin{equation}
\frac{n}{2} = -\frac{1}{M} \frac{\partial F_{B,MF} }{\partial \mu_+},
\end{equation}
\begin{equation}
\frac{n}{2} = -\frac{1}{M} \frac{\partial F_{B,MF} }{M \partial \mu_-}.
\end{equation}
 
   Evaluating the trace in $F_{B,MF}$ yields:
\begin{eqnarray}
F_{B,MF}= \frac{1}{4} \sum_{x,x'} \Phi^* (x) (t_B ^{-1})_{x,x'} \Phi(x')
\\ \nonumber
- \frac{M \mu_+}{2} -T\sum_{x} \ln \sum_{i=1}^4 e^{\lambda_i } ,  
\end{eqnarray}
with
\begin{equation}
\lambda_{1,2}= \pm \sqrt{\mu_+ ^2 + |\Phi(x)|^2}/2T , 
\end{equation}
and
\begin{equation}
\lambda_{3,4} = [2\mu_- -U \pm \sqrt{(U-\mu_+) ^2 + |\Phi(x) |^2}]/2T. 
\end{equation}
$M$ is the number of lattice sites.
At $T=0$ one can therefore neglect the terms containing
$\lambda_2$ and $\lambda_4$, to find
\begin{equation}
\frac{1}{4t_B}= \frac{( (\mu_+ ^2 + |\Phi|^2)^{-1/2} e^{\lambda_1} +
((U-\mu_+) ^2 + |\Phi|^2)^{-1/2} e^{\lambda_3} }{ e^{\lambda_1}+e^{\lambda_3}},
\end{equation}
\begin{equation}
\frac{n}{2} = \frac{1}{2}[ 1+ \frac{ \frac{\mu_+}{\sqrt{\mu_+ ^2 + |\Phi|^2}}
e^{\lambda_1} + \frac{\mu_+ -U }{\sqrt{(U-\mu_+) ^2 + |\Phi|^2} }  e^{\lambda_3}}{
e^{\lambda_1}+e^{\lambda_3}} ] , 
\end{equation}
\begin{equation}
\frac{n}{2} = \frac{e^{\lambda_3} }{e^{\lambda_1}+e^{\lambda_3}}, 
\end{equation}
where we have also assumed a uniform $\Phi(x)=\Phi\neq 0$.

 The mean-field Eqs. (30)-(32) can be solved for any $U$, but we focus here
 only on the limits of large and small interaction. When $U=0$, one finds
  \begin{equation}
  K= \frac{\Phi^2}{4t_B}= n (2-n) 4t_B + O(U). 
  \end{equation}
 The phase stiffness of the non-interacting ($U=0$) weakly coupled dSC
 is a monotonically decreasing function of doping $x=1-n$, behaving in
 essentially the same way as the saddle-point  amplitude
 $\Delta_0$, except for being much larger. This is the familiar
 non-correlated BCS limit. In the strongly interacting case $U/t_B \gg 1$,
 on the other hand, 
 \begin{equation}
 \frac{K}{4t_B} = 2n (1-n)+ 2n^2 (1-n) \frac{4t_B }{U} +
 O[(\frac{t_B }{U})^2].
 \end{equation}
 As $n\rightarrow 0$ the superfluid density, of course, vanishes
 for any $U$. For strong interaction, however, as $n\rightarrow 1$ ,
 $K \rightarrow  0$ for $U/4t_B >2$.
 The order parameter is a decreasing function of $U$, and
 its evolution with interaction is presented on Fig. 2.

 Physical mechanism behind the suppression of the phase stiffness
 near half filling is simple: although the number of bosons per site 
 near $n=1$ is only $\sim 1/2$, the remaining sites are largely occupied
 with spin-down particles. For $U/4t_b > 2$ and right at half
 filling this completely blocks the motion of bosons
 and brings the stiffness to zero. For a small $U$, however,
 it is advantageous for bosons to be in the superfluid state even
 at half-filling, since there is only a small energy cost to hop through
 a site occupied with a spin down electron. At $U=0$, of course, spin-down
 electrons are (almost) invisible to bosons (with $V_{+-}$ neglected),
 and the stiffness at half filling is actually the largest.

 \section{Discussion}

 Our main result is that for a strong repulsion $U$, the $T=0$ 
 superfluid density becomes strongly
 suppressed near half filling, and in fact vanishes right at for large 
 $U$. This is demonstrated for a weakly-coupled superconductor,
 in which the electrons constituting a Cooper pair may be considered essentially 
 uncorrelated in real space. The gist of the method is the introduction
 of two separate density fields for spin up and down electrons, only
 one of which is grouped together with the superconducting phase into a bosonic
 variable, while the other provides the interacting fluctuating
 background. Vanishing of the phase stiffness towards half filling
 is then obtained without invoking any competing order parameters or the 
 real-space pairing.

 That the found behavior of the superfluid density
 implies a pseudogap behavior and the
 nonmonotonic superconducting $T_c$ with doping is qualitatively
 seen as follows.
 In principle, the superconducting $T_c$ is 
 determined by the lower of two characteristic temperatures, the mean-field
 $T_{BCS} \propto \Delta_0$, and $T_\theta \propto K $ at which
 the phase coherence would dissapear even with the amplitude
 $\Delta_0$ held fixed. Since the characteristic energy scale for
 $T_\theta$ is
 the hopping parameter $t$, and for a weakly coupled superconductor
 $\Delta_0 \ll t$, for a weak interaction $U$ $T_\theta \gg T_{BCS}$, and
 it is the $T_{BCS}$ that determines $T_c$ \cite{emery}.
 For strong $U$ however, this will remain 
 true only above some doping $x_{opt}$. For the parameters used
 in Fig. 1 $x_{opt} < 0.5$, since at $x\approx 0.5$ the gap becomes rather 
 small while the superfluid density there peaks. For $x\ll x_{opt}$
 $T_\theta \ll T_{BCS}$
 and it is $T_\theta $ that will determine $T_c$. One can still see the
 remnant of $T_{BCS}$ as the crossover temperature where the density
 of states becomes suppressed, just as it would be in a true superconductor.
 This divorce of $T_{BCS}$ and the true superconducting $T_c$ is an
 unavoidable consequence of strong repulsion, and may be
 taken as an operating definition of a strongly correlated d-wave
 superconductor.

     Our $T=0$ calculation has another interesting consequence worth
mentioning. While a strong $U$ implies the existence of the optimal doping
$x_{opt}$ at which $T_{c}$ peaks, at $T=0$ there is nothing particular 
happening at that doping at all. In fact, as doping is increased
above $x_{opt}$, $T=0$ superfluid density should continue to increase.
This is precisely what is generally observed in high temperature
superconductors \cite{schneider}. Experimentally, the
increase of superfluid density is continued until a higher doping
$x_{opt}'$, above which the superfluid density
again starts decreasing. We believe this effect is likely to be due
to disorder, about which few words therefore need to be said next.

   We found that without any disorder, and with all competing orders
excluded, $T=0$ superfluid density in a strongly correlated dSC is
suppressed near half filling, but always remains finite
at finite doping. Experimentally
on the other hand, it appears that the superfluidity in high temperature 
superconductors is already lost below some critical doping
$x_u \approx 0.05$. The addition of disorder to our bosonic description of
quantum phase fluctuations in underdoped regime would be expected to
lead to precisely such a behavior, as suggested 
by the studies of superfluid-insulator transition \cite{fisher}.
Recalling that electrons, and consequently our effective bosons, in reality
 interact via long-range
Coulomb interaction suggests the universality class of two-dimensional
 dirty bosons with $z=1$, and $\nu\approx 1$ \cite{herbut1},
 in accord with a 
 number of different experiments in underdoped cuprates \cite{singer}.
 In the overdoped regime,
 on the other hand, $\Delta_0$ becomes small and the pair
 breaking effect of disorder
 will bring it, together with $T_c$ and the superfluid density,
 to zero at another critical  doping $x_o$ \cite{herbut2}.
 The precise mechanism of this suppression seems 
 also in agreement with the present, albeit somewhat limited, experimental
 results on overdoped cuprates \cite{schneider}. This would also
 explain why $T=0$ superfluid density starts to decrease for
 $x>x_{opt}'$, where it would still be increasing according
 to our picture, if there was no disorder.

 Although in the present work we were concerned with $T=0$, our picture 
of  the quantum superconductor-insulator transition
 suggests that the universality class of the transition at $T=T_c$ in
 underdoped regime will be of Berezinskii-Kosterlitz-Thouless (BKT) type.
 Indeed, under the assumption of a finite superconducting gap, 
 vortex unbinding becomes the only known mechanism for the loss
 of phase coherence in two dimensions. Possible BKT nature of the
 transition is supported
 by the observed large Nernst effect above $T_c$ in underdoped cuprates
 \cite{ong}, and the measurements of the microwave conductivity \cite{corson}.
 It was argued recently \cite{lee} however, that having the
 BKT transition well below $T^*$ set by a large amplitude
 $\Delta_0$ requires a competing order parameter. The argument is as
 follows: the condensation energy per unit area
 of the BCS superconductor is $\sim {\cal N} \Delta_0 ^2$, where
 ${\cal N}$ is the density of states at the Fermi level. Since the
 coherence length is $v_F/\Delta_0$ one finds that the core energy
 of a vortex is $\sim e_F$, the Fermi energy, which is much larger than
 the BCS superconducting $T_c$. This
 is why the superconducting transition of a weakly coupled superconductor
 is essentially mean-field in character, with an unobservably narrow critical
 region. It seems then that a competing order developing
 in the vortex core is necessary
 to bring the core energy down, so that vortices may proliferate
 at a low $T_c$. That this, however, does not necessarily follow, one may
 see by realizing that the above core energy ($\sim e_F$) is nothing
 but the bare stiffness for the phase fluctuations. The role of the
 competing order parameter would then be to reduce this stiffness with
 underdoping. But as we argued, this seems perfectly possible without
 any ordering, simply from the repulsion which suppresses phase
 coherence even at $T=0$. As long as the $T=0$
 stiffness is small, the BKT transition temperature will also be small
 \cite{case}. The problem with the argument in favor of the competing order
 is that it ignores possible quantum fluctuations
 arising from interactions, which are on the other hand,
 the main point of the present work.

 \section{Acknowledgement}

This work was supported by NSERC of Canada and the Research Corporation.
The author also thanks Matthew Case for useful discussions.

\section{Appendix: derivation of the bosonic mean-field energy}

 Here we present the details of the derivation of the Eq. (23) for the
 bosonic mean-field energy. Within the local approximation discussed
 right above Eq. (22), the partition function for the effective
 bosonic system becomes
 \begin{equation}
 Z_B =\int D[\Psi^*,\Psi] D\rho_- e^{-\int_0 ^\beta d\tau L_B }, 
 \end{equation}
 with the bosonic Lagrangian
\begin{eqnarray}
L_B (\tau) = \sum_x \Psi^* (x,\tau) \partial_\tau \Psi(x,\tau) -\\ \nonumber
t_B \sum_{\langle x,x'\rangle} \Psi^* (x,\tau) \Psi (x', \tau)\\ \nonumber
+U' \sum_{x} |\Psi(x,\tau)|^2 (|\Psi(x,\tau)|^2 -1) \\ \nonumber
+  U \sum_{x} |\Psi(x,\tau)|^2 \rho_- (x' ,\tau) \\ \nonumber 
- \sum_{x} (\mu_+ |\Psi(x,\tau)|^2 + \mu_- \rho_- (x,\tau)) \\ \nonumber 
+U' \sum_{x}\rho_- (x,\tau) (\rho_- (x,\tau)-1),
 \end{eqnarray}
where we have written the hard core interaction $U'$
($U' \rightarrow \infty $) explicitly. This way one can still
write the standard coherent state representation of the
partition function. One can then derive Eq. (23) as follows. First,
decouple the hopping term by introducing
the Hubbard-Stratonovich complex field $\Phi(x,\tau)$:
\begin{eqnarray}
e^{\int d\tau
t_B \sum_{\langle x,x'\rangle} \Psi^* (x,\tau) \Psi (x', \tau)}\propto
 \int D[\Phi^*, \Phi] \\ \nonumber
e^{-\int d\tau [ \sum_{x,x'} \Phi^* (x,\tau)
(t_B ^{-1} )_{x,x'} \Phi (x,\tau) + \sum_x \Phi(x,\tau) \Psi^* (x,\tau)
+c.c. ] }.
\end{eqnarray}
Next, calculate the partition function in the saddle point approximation.
Assuming that at the saddle point $\Phi(x,\tau)=\Phi(x)$ and
$\rho_- (x,\tau) = \rho_- (x)$, i. e. the {\it static} configuration,
in the hard-core limit $U'\rightarrow \infty$ the partition function becomes
\begin{equation}
Z_{B,MF}= e^{-\frac{1}{T} \sum_{\langle x,x'\rangle} \Phi^* (x) (t_B^{-1})
_{x,x'} \Phi (x')} Tr e^{-\frac{\hat{H}_{MF}}{T}}, 
\end{equation}
where the mean-field Hamiltonian is
\begin{eqnarray}
\hat{H}_{MF}= \sum_x ( \Phi(x) \hat{b}^\dagger(x) + \Phi^* (x) \hat{b}
(x) ) - \\ \nonumber
\sum_x (\mu_+ \hat{b}^\dagger(x) \hat{b}(x) + 
\mu_- \rho_-(x) ) +U\sum_x \hat{b}^\dagger(x) \hat{b}(x) \rho_-(x),
\end{eqnarray}
and the trace is to be evaluated over the four states as
described below Eq. (23).
Since $\hat{H}_{MF}$ is local it factorizes into a product over the lattice  
sites, and thus the trace can be easily computed.
Finally, rescaling $\Phi\rightarrow \Phi/2$
leads to the mean-field free energy as displayed in Eq. (23).


\begin{thebibliography}{99}
\bibitem{remark1} $V_{++}$ and $V_{+-}$ are singular at the origin
only in two dimensions and at zero temperature.
\bibitem{comment} Having a large but finite on-site interaction
between same-spin particles does not change
our results in an essential way, but only makes the derivation that follows
more involved.
\end{thebibliography}
\end{document}